\documentclass{appolb}

\begin{document}
\title{Shell Effects in Bubble Nuclei, Atomic Clusters, and Inhomogeneous Neutron Matter
\thanks{Presented at XXXV ZAKOPANE SCHOOL OF PHYSICS,
Zakopane, Poland, 5 - 13 September 2000.}%
}
\author{Aurel Bulgac$^{1}$ and Piotr Magierski$^{2}$
\address{$^{1}$Department of Physics,  University of Washington,
Seattle, WA 98195--1560, USA \\
 $^{2}$Institute of Physics, Warsaw University of Technology, Warsaw, POLAND}
}

\maketitle
\begin{abstract}

We analyze the character of the shell effects/Casimir energy in
inhomogeneous fermion systems.  We estimate magnitude of the shell
effects and discuss their dependence on a number of physical
parameters (geometry, fermion density, temperature).

\end{abstract}
\PACS{ 21.10.Dr, 21.65.+f, 97.60.Jd, 67.55.Lf}

\bigskip

The problem of relative arrangement of impurities inside a many
fermion system is determined solely by the quantum effects, as the
volume, surface or curvature terms in the liquid drop expansion of the
total energy are not affected when impurities are moved around (if
Coulomb effects are irrelevant).  The properties of the quantum
systems that contain impurities of various shapes have been studied in
the case of the Bose--Einstein condensate \cite{chin}.  There, due to
the fact that all the particles have the same single particle wave
function, the behavior of the impurity is rather obvious and one can
easily show that it will be expelled from the condensate \cite{chin}.
In the case of fermion systems the many--body wave function has a
complex character due to the Pauli principle and the answer is not
obvious.

The additional energy associated with the relative arrangement of
impurities can be termed shell correction energy.  In fact, there is
no well established terminology for the energy corrections we are
considering here, even though the problem has been addressed before to
some extent by other authors. In the case of finite systems, the
energy difference between the true binding energy and the liquid drop
energy of a given system is typically referred to as shell correction
energy. In field theory a somewhat similar energy appears, due to
various fluctuation induced effects and it is referred to as the
Casimir energy \cite{casimir}:
\begin{equation}
E_{Casimir} = \int _{-\infty }^\infty d\varepsilon
\varepsilon [g(\varepsilon ,{\bf l} )-g_0(\varepsilon )],
\end{equation}
where $g_0 (\varepsilon )$ is the density of states per unit volume
for the fields in the absence of any objects, $g (\varepsilon ,{\bf
l})$ is the density of states per unit volume in the presence of some
inhomogeneities and ${\bf l}$ is an ensemble of geometrical parameters
describing these objects and their relative geometrical
arrangement. Thus the Casimir energy can be thought of as a measure of
the fluctuations induced in the energy spectrum in the presence of
various ''obstacles''.

There are a number of situations when such inhomogeneities can be formed.
In the nuclear physics it was suggested a long time ago that very
heavy nuclei will tend to develop a hole inside in order to minimize
the Coulomb energy \cite{wheeler,pomorski,bender} and similar objects
can be created in the case of atomic clusters \cite{dietrich}.

The creation of voids is also predicted to happen in the nuclear
matter at subnuclear densities. Apparently, an agreement has been
reached in literature concerning the existence of the following chain
of phase transitions as the density increases: nuclei $\rightarrow$
rods $\rightarrow$ plates $\rightarrow$ tubes $\rightarrow$ bubbles
$\rightarrow$ uniform nuclear matter \cite{pethick}.  In these cases,
fermions reside in a rather unusual mean--field, which is much deeper
for bound than for unbound nucleons.  Since the amplitude of the wave
function in the semiclassical limit is proportional to the inverse
square root of the local momentum, the single particle wave functions
for the unbound states will have a small amplitude over the deep
well. Hence the deep well will act almost like a hard wall (in most
situations).  For the same reasons the halo nuclei \cite{austin} can
be thought of as some kind of bubbles as well.

Let us consider first the problem of positioning of a spherical bubble
inside a finite spherical Fermi system.  The total energy can be
expressed in the form:
\begin{equation}
E(N)=E_{LD}(N)+E_{shell}(N) = e_vN +e_sN^{2/3}+e_cN^{1/3} + E_{shell}(N),
\label{eq:liq}
\end{equation}
where $E_{LD}$ is the smooth liquid drop part of the total energy and
$E_{shell}$ is the quantum shell correction contribution to the total
energy. As we mentioned above, once the bubble is formed its
displacement will not affect either the volume, surface or curvature
terms in the liquid drop expansion. Hence, classically moving bubble
off--center costs no energy, if Coulomb energy is left aside for the
moment. The closer investigation of the shell correction energy shows
that it depends strongly on the position of the bubble.  The most
pronounced shell effects are predicted for the spherically symmetric
system (bubble in the center). In this case the system is classically
integrable and thus its quantum--mechanical counterpart will have
large gaps in the spectra resulting in a large amplitude of the shell
correction energy. The rapid fluctuation of the shell energy as a
function of the fermion density indicate that for some numbers of
fermions it would be energetically more favourable to expell the
bubble off--center. For a finite eccentricity the system is
chaotic and the more the bubble is shifted from the center
the larger part of phase-space is occupied by chaotic
trajectories. Nevertheless, the shell effects are still strong.
One can show that the shortest periodic orbit
determines the gross structure of the shell energy when the
bubble is close to the surface \cite{yu,bul}.

The problem of two or more objects immersed in an infinite Fermi
system has a similar character.  In an infinite system the presence of
impurities results in an appearance of resonances, which contribute to
the shell correction energy. In order to better appreciate the nature
of the problem, let us consider the situation when two identical
spherical bubbles have been formed in an otherwise homogeneous Fermi
system. We shall ignore here the possible Coulomb interaction, as its
main contribution is to the smooth part of the total energy of the
system.  In the semiclassical approach, which is justified for the
''sizeable'' bubbles (i.e. when the Fermi wavelength is small
comparing to the size of the bubble), the shell correction energy is
determined by the periodic orbits in the system.  In the case of two
spherical bubbles there exist only one such trajectory (with
repetitions) which gives rise to the interaction energy between
bubbles. It is the hyperbolic orbit lying in the line connecting the
bubbles centers and characterized by the Lyapunov exponent: $\lambda =
2 \ln \left [ 1 +\frac{d}{R} + \sqrt{ \frac{d}{R}\left ( \frac{d
}{R}+2 \right ) } \right ] $, where $R$ is the radius of the bubble
and $d$ is the distance between their centers. The interaction energy
between the two bubbles due to the existence of this periodic orbit
reads:
\begin{equation}
E_{shell}=
\frac{\hbar^{2}k_F^2}{2m}
\frac{1}{4\pi (k_F d)^2} \sum_{n=1}^{\infty}
\frac{ [ 2 n k_{F} d
 \cos (2 n k_{F} d) -
   \sin (2 n k_{F} d) ]}
{n^{3} \sinh ^{2}(n\lambda /2 )},
\end{equation}
where $k_{F}$ denotes the Fermi momentum and $m$ the mass. Similar
arguments can be presented for other shapes, e.g.
cylinders or plates. One finds that at large separations the
interaction energy oscillates as well but it decays as $\propto
1/d^{5/2}$ in case of cylinders and as $\propto 1/d^2$ in the case of
plates.

Hence it is clearly seen that there are some preferable arrangements
in the systems of two impurities. The interaction between them depends
mostly on their shapes and the geometry of the mutual arrangement. Our
results show that shell effects associated with the appearance of
inhomogeneities in the neutron matter at subnuclear densities may
strongly influence the phase transition pattern in the neutron star
crust since the shell effects for various ''nuclear lattices'' are of
the same order as the differences of the total energy between
phases\cite{bul1}.

We suspect that there are a lot of other effects, which might be
relevant. We did not consider periodic orbits bouncing between three
or more objects. An orbit bouncing between two bodies leads to a
pairwise interaction. Orbits bouncing between three or more bodies
would lead to many body interactions, which however turn out to be of
lesser importance \cite{bmw}.  We have also considered only perfectly
smooth objects. If one allows for some degree of corrugation of these
surfaces, many more periodic orbits are likely to appear and that
would lead to even more complicated interactions and more complicated
interference patterns. If temperature has a simple to predict
qualitative effect on the shell correction energy, the role of pairing
effects is not that transparent and deserves further investigation.

We appreciate the financial support of DOE and 
of the Polish Committee for Scientific Research under Contract No. 2 P03B 040 14.

\end{document}